\documentclass[twoside,12pt]{article}
\usepackage{graphicx}
\usepackage{amsmath}
\usepackage{amsfonts}

\usepackage[round,authoryear,sort&compress]{natbib}

\title{Adapting CBPP Platforms for Instructional Use}

\author{R. Milson (Dalhousie U.)\thanks{Supported by Dalhousie's
    Centre for Learning and Teaching, and N.S.E.R.C., Canada.
  Email address: \texttt{rmilson@dal.ca}} \and A. Krowne
  (Emory U.)\thanks{Supported by the Digital Library Research
    Laboratory, Virginia Tech, on NSF DUE-0121679.  Email address: 
    \texttt{akrowne@emory.edu}} }

\newcommand{\noo}{No\"osphere}
\begin{document}
\maketitle

\begin{abstract}
Commons based peer-production (CBPP) is the de-centralized,
net-based approach to the creation and dissemination of information resources.
Underlying every CBPP system is a virtual community brought together
by an internet tool (such as a web site) and structured by a specific
collaboration protocol.  In this talk we will argue that the value of
such platforms can be leveraged by adapting them for pedagogical
purposes.

We report on one such recent adaptation. The \noo\ system is
a web-based collaboration environment that underlies the popular
Planetmath website, a collaboratively written encyclopedia of
mathematics licensed under the GNU Free Documentation License (FDL).  Recently, the system was used
to host a graduate-level mathematics course at Dalhousie University,
in Halifax, Canada.  The course consisted of regular lectures and
assignment problems.  The students in the course collaborated on a set
of course notes, encapsulating the lecture content and giving
solutions of assigned problems.  The successful outcome of this
experiment demonstrated that a dedicated \noo\ system is well 
suited for classroom applications.  We argue that this ``proof of concept'' experience
also strongly suggests that every successful CBPP platform possesses 
latent pedagogical value.
\end{abstract}

\section{Introduction}

\subsection{Background and motivation}

The capacity of communications networks to create value is well
recognized \citep{Metcalfe}.  There is a theoretical argument that
internet value creation is an even more dramatic process, because it
is dominated by exponential rather than polynomial scaling effects
\citep{Reed}.  To put it another way, the internet engenders powerful
emergent phenomena, because every potential group with a shared
interest can interact, collaborate, and create intellectual value 
through internet (and especially WWW) software applications.

Thus, with the advent of powerful search and indexing technologies,
the world wide web is evolving into a ubiquitous reference
resource \citep{BernersLee}.  The network
transforms the disconnected efforts of millions of web page authors
into something of practical value.  Another noteworthy project is Wikipedia
\citep{Wales}, a knowledge-oriented virtual community that
successfully employs the wiki collaboration protocol \citep{Leuf}
to unite the efforts of thousands of volunteers
around the scholarly goal of a public domain encyclopedia \citep{Kantor}.

In both of the above examples, the underlying process lacks
explicit organization and is non-hierarchical. In both cases the value is
governed by an emergent phenomenon: the value of the whole is
significantly greater than the sum of the individual parts.  A recent
economics-based theory attempts to explain such emergent value
phenomena as instances of \emph{commons-based peer production}
(CBPP), an idealized mode of production that is complementary to firms
and markets, and one that manifests naturally on the internet
\citep{Benkler}.  However, economic theory is insufficient to fully
understand and exploit the complex, emergent phenomena that underly
internet value creation \citep{Iannacci}.  The study of the internet is inherently
cross-disciplinary; no one discipline, or even a blend of two will
suffice.

In the present article we report on and discuss a recent adaptation of
\noo\ \citep{Krowne2003a}, a web platform for
mathematics collaboration, for the purpose of teaching a graduate
course in mathematics.  A convenient categorizing label for our
project is \emph{computer supported collaborative learning} (CSCL),
a field that brings together perspectives from cognitive science,
computer and information science, education, and philosophy \citep{Stahl2006}.
Our thesis is inherently cross-disciplinary. We argue
that CBPP, the phenomenon of internet value creation, crosses over
naturally into the world of CSCL.  We argue that the infrastructure of
collaborative, knowledge-related projects, like Wikipedia and
\noo, can be leveraged to yield concrete educational assets.

This value stems in large part from the inherent unity and
collaborative nature of the scholarly enterprise. A context that
fosters the formation of communities which acquire, organize, generate, synthesize, and
transmit knowledge will also be a context where learning and pedagogy
are of central importance.  These qualities naturally lead us to the
concept of a \emph{digital library}.  Traditionally, libraries have been the
cornerstone of scholarship, providing a space for both research and learning,
and other,  more intangible benefits.  It would therefore be surprising if 
emergent collaboration phenomena and educational scenarios
did not play a role in the evolution of the digital library \citep{Robertson}.

\subsection{Re-conceptualizing the digital library}  

The concept of a digital library is a natural outgrowth of the
development of modern, network-oriented information technology.
Information, once digitally encoded, can be stored electronically and
distributed over the internet.  Physical and geographical barriers
disappear.  There are no limits to the size of the library.  It's
contents are potentially available to everyone, everywhere, all the 
time.

The word \emph{library} carries with it connotations of a nearly
static archive, one where the primary information-related activity are
storage, classification and retrieval.  The shift of information
content from the physical to the digital realm undermines this
traditional conceptualization \citep{Levy}.  Various recent
internet-focused developments--- powerful and ubiquitous search
engines, virtual communities and the free culture movement, to name
just a few--- challenge us to move beyond the simple notion of an
``electronic traditional library,'' and to embrace benefits beyond the
elimination of space and scarcity concerns.

Older information technologies, such as paper, foster a dichotomy
between \emph{information} and \emph{knowledge}.  The latter is the more
dynamic concept; knowledge implies research, dissemination, debate, synthesis,
activation, history and evolution.  As well, knowledge cannot be conceived
as something separate from people; knowledge implies a community of
scholars, teachers, learners, and practitioners \citep{Ehrlich}.

Therefore, the digital library concept needs to evolve to more fully realize
the potential of the underlying network technology and software technology.
New library tools and modalities that address collaboration, superimposed
information, knowledge creation, and education will have to be developed
\citep{Delcambre, Krowne2003b, McRobbie, Frumkin}.

\subsection{CBPP}

In this regard, \emph{commons based peer production} (CBPP) shapes up to become
a key phenomenon in the digitally mediated transition from \emph{information}
to \emph{knowledge}.  Internet-based CBPP has its origins in the open-source
software movement, a collaborative, extra-commercial process of software
creation\footnote{This is not imply that open source software is without
commercial value.  Rather, the process of creation is governed by something
other than a simple exchange of money for software end products.}. The
existence of numerous successful internet projects, Wikipedia and
Project Gutenberg/Distributed Proofreaders \citep{Lebert}, to cite just
two examples, indicate that the phenomenon of collaborative internet
value creation has pertinence well beyond generating software programs.

With peer production on the Internet, distributed ensembles of people
share open production of complex products and services--- generally
for no financial compensation.  While the idea of non-market,
non-corporate production is not new (science has traditionally worked
this way), large-scale, decentralized, sustained, open production by
diverse groups of peers is a new phenomenon: a development that has
been enabled and encouraged by the confluence of computers, networking
and the information economy.  
This form of non-market, internet-based
peer production has been applied to create a wide variety of
significant knowledge assets \citep{Galiel}.

The impact of a knowledge-centric community like Wikipedia on the
digital library landscape cannot be ignored.  Neither should the enormous 
productive leverage of a project like Distributed Proofreaders. 
Therefore, it makes good sense (for both practical and idealistic
reasons) to expand the ``digital library'' concept to incorporate an
internet-based CBPP aspect.  

PlanetMath \citep{Krowne2001} is another CBPP project, of
special connection to our study.  Planetmath is a collaboratively
written encyclopedia of mathematics licensed under the GNU Free Documentation
License (FDL), and implemented using the \noo\ system.  The
PlanetMath project is an instance of CBPP; the aim is to create a
community-oriented, web-based repository for mathematical knowledge.
The project attracts a diverse and international body of participants.
These people are students and members of the wider public with an
interest in mathematics, graduate students pursuing advanced
mathematics degrees, professional mathematicians who make their
living by practicing or teaching mathematics classes and by conducting
mathematics research.  Planetmath and \noo\ also have an extended role as a
testbed for research and development in semantic extraction, digital
information exchange, and collaborative authority models \citep{Krowne2004}.  

\subsection{Academia, instruction, and engagement} 

Academic communities are concerned with knowledge in all its
manifestations; both the information and community-related aspects are
important.  Certainly, instruction and the teacher-learner
relationship are central academic concerns.  

Instruction can be conceptualized as a structured interaction between
senior and junior members of a knowledge community.  The
instructor is more than just a particular medium for the storage and
transmission of information.  Rather, for the student, the lecture
hall is a portal to the community of knowledge \citep{Clancey}. Let us
use the term \emph{engagement} to describe the process of active
student participation and scholarly development \citep{Stahl2005}.

In addition to the immediate goals of any particular course of
academic instruction, there is, in the teacher-student relationship,
an implicit invitation to ``do as we do''; to join the community, and
to become involved in knowledge-related activities.  Pedagogical
structures:  exercises,  discussions, individual and group
projects,  examinations and other assessment modalities, are the
devices of guided scholarship.  Engagement, rather than skill-set
and information ``download'' is the deeper goal of academic instruction.
The ultimate measure of success is the metamorphosis of the \emph{student},
an individual at the outset capable and interested only in passive,
assisted knowledge activities, into the \emph{scholar}, an individual
engaged in independent knowledge activities.  

It is worth briefly examining the critical elements of scholarship.  Of paramount importance
is that for scholar, no ``oracle'' exists to provide the answer to a research
question.  Peers can provide critique but not guaranteed answers.  The scholar
also lacks a roadmap towards a solution, and must prioritize his/her efforts,
evaluate the intellectual contributions of others, and act upon their own
judgments.  This is the universal situation of the scholar, and it is
utterly different from the environment of the formal student.  While
attempts are made to deliberately teach students many of the tools upon
which scholars rely, the aims and trajectory of classroom activities are by
definition preset.  Thus, the characteristics of the true scholarly
environment induce a sharp division of students who have meaningfully become
scholars from those who have merely learned to regurgitate information with
relative success.  

CBPP projects like Wikipedia and \noo\ possess a remarkable capacity for
fostering engagement in scholarly activity.  We suggest that it is reasonable to
tap such free-culture phenomena for the purposes of academic instruction.
Indeed, nothing could be more natural, because of the inherent compatibility
between academic and free-culture goals and values \footnote{The open access
movement illustrates this nicely \citep{Suber}.}.  Let us make a sketch of
how such an evolution can take place.

A re-conceptualized, more dynamic and community-oriented digital library is a
natural context for both public domain knowledge activity and for pedagogical
efforts that involve students in online knowledge activities.  Such activities
should include not just information retrieval, but collaborative knowledge
creation and organization \citep{Brown}.  The physical community of the
classroom can be extended to the network.  The same community and collaboration
tools and technologies that enable CBPP projects can be used to create a
virtual space in which the participating students can carry out
knowledge-related activities, albeit in an assisted and structured fashion.

We hypothesize that such an approach can lead to a heightened level
engagement, because of the subtle but important shift of emphasis from
``I will teach, you will learn'' to ``let us collaborate on a
knowledge project''.  The change of attitude is natural and desirable
from an academic point of view, but is difficult to implement using
traditional classroom methods and technologies.

Our hypothesis is that adoption of CBPP technologies into an instructional
setting will facilitate just such a shift of emphasis.  The student goal-set
and motivations will be enriched by incorporating a network-based,
collaborative aspect into the classroom experience.  At one level, the
instruction process can proceed in the traditional manner: the teacher
guides the students through a fixed syllabus, assigns tasks, and performs
evaluation.  However, since the setting is now a ``research library'' as
well as the classroom,  since the medium of interaction includes a
virtual collaboration environment, and since the goal-set includes the
incorporation of individual efforts into a digitally encoded body of
knowledge, the end result will manifest as a collaboration between all
involved.  Such a process should lead to heightened levels of student
engagement.

\section{A Trial of \noo\ as a platform for collaborative instruction}

\subsection{Test scenario and goals}

In the Winter of 2003, the \noo\ system was used to host \textbf{Math
  5190: Ordinary Differential Equations}, a graduate mathematics
course at Dalhousie University, in Halifax, Canada.  One of the
current authors served as course instructor.  A ``tabula rasa'' \noo\
system was set up on a dedicated server.  The primary course goal was
the collaborative creation of a set of course notes, including a
number of worked-out exercises to illustrate the key concepts.
Assessment criteria included the quantity and quality of the online
participation, as well as a more conventional final project.

The course attracted 3 graduate students and an auditor, who in the
coming semester created and organized an online body of knowledge on
the topic of differential equations.  The end result was a 70 page
document containing definitions, theorems, proofs, and examples.  
When taken together, these constitute a mini-treatise on certain aspects 
of the theory of ordinary differential equations.

The trial addressed the following research goals:

\begin{enumerate}
\item Our main hypothesis was that CBPP platforms are suitable for
  advanced mathematics instruction, and that a course structured
  around collaborative principles and online tools can serve and
  advance conventional academic goals.
\item We evaluated the feasibility of deploying \noo\ as a CSCL
  environment.  Experiences with CoWeb \citep{Guzdial2001}, show
  that CSCL-type mathematics courses present special challenges
  related to specialized notation and division of labor issues.
  \noo's \LaTeX{}-based design incorporates the full range of
  advanced mathematical notation.  As well, \noo\ possesses a
  unique authority model and groupware capabilities.  The trial
  examined the capacity of these designs to address the above
  challenges.  In particular, we wanted to compare the patterns of
  student activity in a collaborative, online environment with those
  in a traditional mathematics courses, and to consider the impact on
  student engagement.  Our secondary hypothesis is that student
  engagement benefits from the introduction of CBPP elements.
\item We also considered the impact of a collaborative, online course
  environment on the students' scholarly development.
\end{enumerate}

\subsection{Methodology}

Math 5190 is a one-semester course at Dalhousie University on the theory
and methods of ordinary differential equations.  Such courses,
typically aimed at beginning graduate students and advanced
undergraduates, are offered, with certain variations, by most
mathematics departments in North American universities.

In the Winter of 2003 this course served as a proof-of-concept study
of the \noo\ system in an educational setting
The course included a number of conventional
instructional components: 3 hours/week of lectures, a reading list,
regular meetings with of the instructor with individual students, a
final project, and student presentations.  The core component,
however, was a dedicated website set up as a ``tabula rasa''
\noo\ environment.

The basic unit of content in \noo\ is the \emph{entry}, which any
registered user can create. The entries comprise the main section of
the system, which is called the ``encyclopedia''. This reflects the
general orientation and pedagogical style of the system. 

\noo\ entries consist of title, content, and various metadata. The
entries are interlinked, which means that the text of each entry
contains hyperlinks pointing to other entries where appropriate. The
general intent of this is to provide definitions for each concept
utilized, in an easily navigable fashion. Entries are written in \LaTeX\ 
\citep{Lamport}, which serves as the basis for \noo's mathematics
support in addition to allowing for the expression of general document
formatting. Displayed in rendered form, the mathematical portions of
each entry ``look right'' with a standard browser (with no plug-ins),
a considerable improvement over most other attempts to publish
mathematics to the web to date. This mathematics support makes \noo\ a
good candidate for use in all of the mathematical sciences.

A key feature of \noo\ is the \emph{corrections} system. If any
registered user determines there is a problem with an entry, he or she
can voice concern by filing a correction to that entry. Until
addressed, this correction is displayed when the entry is shown,
ensuring that the critique is ``out in the open''. 

Finally, each entry in \noo\ has an owner, who is initially the person
who created the entry.  An owner has the option of \emph{orphaning} an
entry, or transferring ownership to another user.  Orphaned entries
are flagged by the system and may be \emph{adopted} by any interested
user.

\noo\ has a number of other services that provide direct community
support.

\begin{enumerate}
\item The \emph{requests} service, which functions as a global ``to-do'' list
of content addition for the \noo\ site.  Users can fulfill
particular requests, rendering them inactive, by creating an
appropriate entry.

\item 
The \emph{discussion} service provides threaded, asynchronous
messaging. A discussion can be attached to most of the core objects
of \noo\ . This includes encyclopedia entries, corrections, and
requests.

\item 
\noo's \emph{notification} system keeps members of the community aware of
activity relevant to them through e-mail and a \noo\ system ``inbox''.
Corrections to an entry result in a notice to the entry's owner. A resolved
correction results in a notice to the filer, indicating what action was taken
and why. Similarly, replies to a message posted result in a notice that makes
the initial poster aware of the reply.  An important part of the notification
system is the ability to create configurable \emph{watches}. Watches placed on
any object by any user result in (e-mail or web) notices about events to that
object being sent to the user. 
\end{enumerate}

At the outset, the students were informed that the main course
objective was the collaborative creation of a set of lecture notes
using the online environment.  The instructor's role was to facilitate
and to structure this effort.  As such, the instructor mirrored
lecture topics and contents with \noo\ \emph{request} objects
that enumerated the key definitions, theorems, proofs, and techniques
covered in the lectures.  The students were responsible for filling
these requests by creating the requisite \emph{entries} and
subsequently evolving and improving them based on \emph{corrections}
received from the instructor and fellow classmates.  The students had
to cooperate to decide how to divide the requests and to share the
corresponding workload.

It is well recognized that mathematics instruction is greatly
facilitated by supplementary problems and exercises.  In place of the
conventional system of regular assignments with specific deadlines,
course exercises were presented to the students as illustrative
examples to be included in the collaborative notes.  The instructor,
on a regular basis, created and \emph{orphaned} exercise-type
entries.  The students were then responsible for \emph{adopting} the
entries and furnishing solutions.  Again, students were given the
opportunity to evolve and improve their solutions through interactions
with instructor and classmates.  \emph{As such, an incorrect solution did
not necessarily result in a poorer evaluation, but rather served as an
additional learning opportunity} in the context of \noo's system
of corrections. Students had the opportunity to continuously improve
their entries up to the course termination deadline.

The collaborative, online aspect of student progress was assessed
according to the number of owned entries, and according to the extent
the entries were developed.  At the termination of the course, a score
of 1,2, or 3 was assigned to each student entry according to the
following criteria:

\begin{itemize}
\item Degree of participation was measured by the number of
  filled requests, and adopted exercises.  An adopted entry with even
  a minimal amount of content was assigned a score of 1.
\item A reasonably well developed entry with unresolved
  corrections was assigned a score of 2.
\item A correct, well written entry with no outstanding corrections
  was assigned a score of 3.
\end{itemize}

The instructor issued corrections in response to student errors, and
to suggest improvements to the mathematical content and presentation
format.

Course assessment did not include an examination component.  Rather,
an assessment of scholarly development was based on a final project,
which was implemented conventionally, and involved both an oral
presentation and a written report.  With input from the instructor,
students selected a relevant topic\footnote{The 3 registered students
  chose the following topics: convergence of iterative integral
  solutions, predator-prey models, differential equation modeling of
  guerrilla vs. conventional warfare.}, delivered a classroom
presentation, and submitted a written report.  The project component
played a particularly important role in the trial, providing a measure
of student progress independent of the online activity.

\subsection{Results data}

By the end of the course, the 3 registered participants, all
first year MSc students, had created a total of 122 entries.  The
entry totals and the corresponding scores (see above) are displayed
in Table \ref{tab:entries}.  A score of 0 indicates an entry with
non-existent or negligible content.  At the conclusion of the course
there were a total 12 unfilled requests and unadopted exercise
problems. A total of 78 corrections were issued\footnote{All but one of these
corrections originated with the instructor.}.

Subsequently, the website contents were converted into document form and
redistributed to the students.  The resulting document spans 74 typeset pages.
The table of contents of the resulting document in displayed in
Appendix~\ref{NotesTOC}.  Some representative entries are shown in
Appendix~\ref{NotesExamples}.

The \noo\ collaboration protocol proved to be very suitable for
student-instructor interactions.  The entry ownership system and
email updates allowed the instructor to easily follow student
progress, and to issue timely feedback in the form of corrections.
With minor adjustments, the \noo\ scoring system proved valuable
as a highly visible indicator of individual participation levels.

\begin{table}[htbp]
  \centering
\begin{tabular}{c|rrrr|r}
& \multicolumn{4}{c|}{Entry score} \\
Student & 0 & 1 & 2 & 3 & Total \\ \hline
1 & 0 & 1 & 10 & 26 & 37 \\
2 & 1 & 2 & 10 & 27 & 39 \\
3 & 3 & 6 & 10 & 16 &32
\end{tabular}
\caption{Student entries and assessment scores.}
\label{tab:entries}
\end{table}

Student-instructor interactions stabilized around the following
cyclical pattern. The instructor delivered lectures and suggested
deadlines for the fulfillment of requests and the adoption of exercise
entries.  This was followed by posted corrections and occasional email
``nags'' and feedback.  As is often the case in conventional courses,
the students functioned as largely passive knowledge agents.  There
was no evidence of direct online collaboration among the students.
Students did not give each other corrections, nor did they use  the
online forums to discuss mathematical content.  Rather, students
reported collaborating in more conventional ways.  They held study
group meetings to discuss course material, and to decide on the
division of labor for their online tasks.

\begin{figure}

\begin{centering}
        \includegraphics[scale=.5,angle=270]{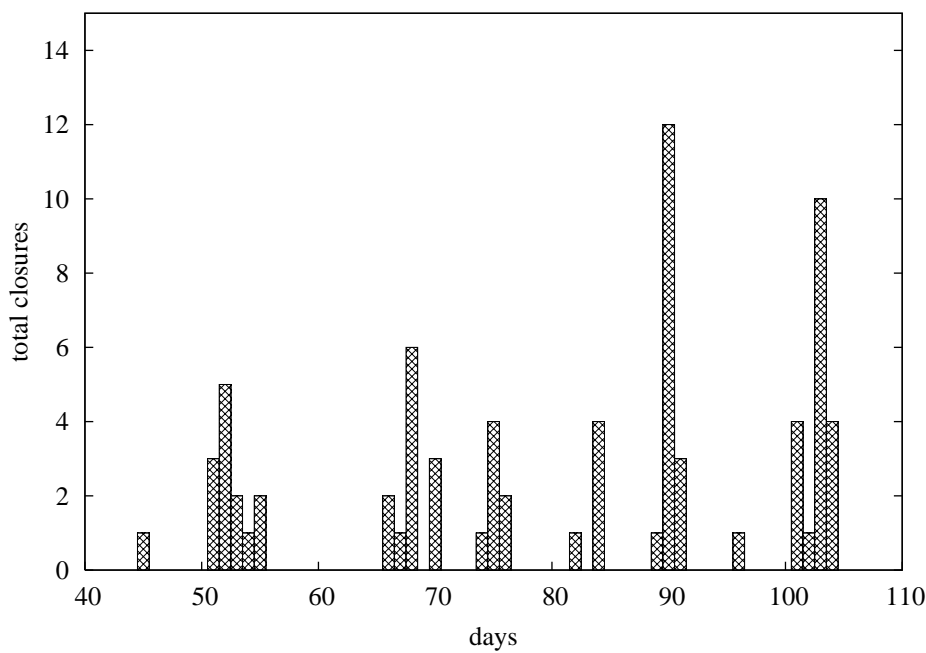}
\end{centering}

\caption{Chart of correction closures by students (with each bar representing a day), revealing the  ``bunching'' effect.}
\label{Closures}
\end{figure}

Student behavior and outlook in the trial was typical for courses at
the beginning graduate level. Students at this level still require
explicit goal structure and assessment criteria, and are often passive
in their approach to the material.  Students in the trial displayed
typical procrastination behaviors, and regarded their participation as
``necessary duty'' to be balanced against time requirements from other
courses and from outside jobs.  As such, their online efforts tended
to occur in bursts of concentrated activity.  An example of this
behavior pattern is visible in Figure~\ref{Closures}, which shows the temporal
distribution of student responses to corrections.

The conventional educational objectives of the course were fulfilled.
The content of the final projects and the website entries, especially
the exercises, provided clear and substantial evidence of progress
toward mastery of the subject matter, and progress in scholarly
development.  Relative to these metrics (exercise solutions and final
projects), progress of the students in the trial was directly
comparable to the progress of students in the same course taught by
the same instructor conventionally in other years.

\subsection{Findings}
Given the limited enrollments and the advanced nature of the material
characteristic of graduate courses, and keeping in mind the natural
variation of student backgrounds and abilities, it is not feasible to
render a  judgment on the relative merit of conventional
pedagogy versus collaborative, online learning.  However, our
observations allow us to make the following points.
\begin{enumerate}
\item Our experience with Math 5190 and \noo\ provides strong
  support for the hypothesis that conventional educational objectives
  can be met by a course based on online learning and CBPP principles.
  Importantly, we found no evidence that the inclusion of a CBPP 
  component diminished or disrupted traditional classroom learning.
  Our outcomes should be reproducible by groups of advanced students
  at other institutions, and with other courses in the mathematics
  curriculum.  To make sense of this claim, however, one must
  incorporate assessment components that can provide an objective
  measure of student progress.
\item The students in the trial readily accepted the mechanics of
  \noo\ and expressed appreciation at being able to do their
  work in an online setting.  Nowadays knowledge of \LaTeX\  is  a
  near-universal prerequisite for the scholarly development of
  mathematics students.  The \LaTeX\  component of \noo\ provided
  our students with a useful opportunity to develop their typesetting
  skills.

  Based on the instructor's observations and communication with the
  students, \noo's protocol of entry adoption and ownership
  allowed the students to exercise control over their participation,
  and thereby facilitated engagement.  The fulfillment of requests and
  the adoption of exercises manifested as an act of commitment on the
  part of a student.  Thus, the authority model allowed the students
  to pursue a division of labor, but in a transparent and principled
  fashion that is usually lacking in conventional courses.

  A potential weakness of this approach is the possibility that an
  overly selective focus on the part of some students may lead to a
  spotty coverage of essential topics.  The instructor has an
  important role to play here, and must encourage students to
  contribute to a variety of course topics.  Such difficulties did not
  visibly manifest in the trial under discussion.  However, without a
  comprehensive final examination it is difficult to discount the
  possibility that some of the students received inadequate exposure
  to some of the topics.

\item The collaboratively compiled course notes are a valuable asset
  that is not readily available in the context of conventional
  instruction.  From the point of view of the students, the document
  is far more than a  transcription of the instructor's lectures.
  In a very real sense, the students are the authors of the
  document. As such, the notes concretely encapsulate their
  learning experience.

  There are a number of benefits to producing such a document.  The
  notes can serve as a source of reference for future work in the
  subject.  Perhaps, more importantly, the very existence of the notes
  embodies a latent, but powerful message about the students' capacity
  for scholarship, and about the nature of the academic enterprise.
  In an important sense, the creation of the notes transforms the asymmetrical
  relationship between instructor and the students into something more
  closely resembling scholarly collaboration.

  There is also the intriguing possibility that collaboratively
  produced course notes can serve as contributions to public domain
  knowledge repositories\footnote{The students in the trial were encouraged
  to convert their course contributions into PlanetMath entries--- though  
  none of them chose to pursue such activity.}.

  The primary responsibility of the course instructor centers around
  the student learning experience.  As such, it would not be
  appropriate to make full scholarly use of the course notes without
  addressing issues of consent and attribution.  Still, it is
  important to provide students with opportunities for independent
  scholarly activity.  If nothing else, the format of the trial made
  the students aware of ongoing CBPP efforts, and served as an
  invitation to contribute to them.
\end{enumerate}

\section{Discussion}

The \noo/Math5190 trial constitutes a limited, proof-of-concept
experiment regarding the application of CBPP tools in an educational,
academic setting.  Though our experiment was a success, the small
scale of the trial limits the inferences we can draw in support for
our hypothesis regarding CBPP and education.  It will be necessary to
subject the hypothesis to further testing: one needs to organize more
CBPP-based courses, involve more students and instructors, employ
control and experimental groups, and to consider diverse academic 
subject material.

One also has to come to grips with the limitations revealed by our
experience.  Collaborative learning methods are not a panacea
for improving student engagement \citep{Guzdial2002}.  Indeed, it
would be useful to undertake a systematic examination of the effects
of CBPP on academic engagement.  Methodologically, the undergraduate
curriculum, with its larger enrollments, may be a more appropriate
setting for such studies.  

Wiki-based courses in the humanities and the social sciences are the
subject of ongoing research and discussion \citep{Boyd}.
Wiki software is widely available, notational demands are lower, and
the wiki interface is easier to learn than the \LaTeX{}-based
\noo.  The ostensible aim of such a course should be a
well-developed body of ``wikified'' content that encapsulates a
subject of interest, and that provides a concrete record of individual
students' participation.  An initial study on this topic \citep{Scharff}
supports the conclusions of our own trial.  It would also be
interesting to study to the effect of such an experience on scholarly
evolution.  To what extent does student exposure to wikis as an
instructional medium encourage contributions to sites like Wikipedia,
or the pursuit of more conventional scholarly publications?

\section{Conclusion}

The joining together of the themes of collaborative education, the
internet, and digital libraries is not a new idea \citep{Roschelle}.  
Rather, the relatively recent emergence of successful CBPP
knowledge projects should be viewed as a timely and complementary
development \citep{Tomek}.  Much of the infrastructure, interface, and
design issues are the same for both contexts.  There is strong common
focus on extraction of semantics, collaboration interfaces, and
educational applications.  We believe the potential for mutual benefit and a
convergence of interests is evident.

In the context of a symposium on Digital Libraries and Free Culture,
it is also appropriate to note the relevance of our hypotheses to
the continuing debate about intellectual property and the public
domain.  Pragmatism and utility are strong arguments for commons-based
knowledge activity.  The impact of the open source and the free
software movements on development of information technology is, at this point,
beyond question.  Likewise, projects like Wikipedia, PlanetMath, and
Distributed Proofreaders are beginning to make a significant contribution
to the intellectual commons.

As is the case with emergent internet value phenomena, the potential
value of such projects is unconstrained and will manifest in unforeseen
ways.  But, this is just one instantiation of the general argument in
support of public domain knowledge and culture \citep{Lessig}.
Synergy and flexibility is the point here, and a \emph{libre} free
project like PlanetMath is good example.  This project
began as a mathematics encyclopedia, then evolved into a groupware
platform and  test-bed for digital library research (\noo),
and is now being used as an educational delivery system.

Academic involvement in CBPP projects allows researchers, librarians,
and educators to exploit the kind of internet value that IT companies
enjoy when they employ open-source software.  Conversely, free-culture
projects benefit from academic attention and investment.  Successful
adaptation of CBPP technologies for academic instruction is a powerful
argument in support of free culture.  However, much work remains to be
done in the cross-disciplinary exploration of CBPP, CSCL, and digital
libraries.


\bibliography{emory}





\newpage 

\appendix

\setlength{\parindent}{0cm}
\section{Math 5190 course notes: table of contents}

\label{NotesTOC}
\bigskip

\begin{center}
\bf Topics in Ordinary Differential Equations\\
  Collaborative course notes
  compiled by the students of Math 5190\\
  Dalhousie University\\
  April 18, 2003  
\end{center}

{\small
\setlength\parskip{-12pt}
\contentsline {section}{\numberline {1}Autonomization}{5}

\contentsline {section}{\numberline {2}Banach fixed point theorem}{5}
\contentsline {section}{\numberline {3}Bernoulli equation}{5}
\contentsline {section}{\numberline {4}Cauchy sequence}{6}
\contentsline {section}{\numberline {5}Cayley-Hamilton theorem}{6}
\contentsline {section}{\numberline {6}Characterization of homogeneous equations}{7}
\contentsline {section}{\numberline {7}Characterization of linear ODEs}{8}
\contentsline {section}{\numberline {8}Characterization of separable equations}{9}
\contentsline {section}{\numberline {9}Characterization of the Bernoulli equation}{10}
\contentsline {section}{\numberline {10}Characterization of trivial symmetries}{10}
\contentsline {section}{\numberline {11}Compact}{11}
\contentsline {section}{\numberline {12}Completeness of a compact subspace}{11}
\contentsline {section}{\numberline {13}Constant coefficient symmetries}{12}
\contentsline {section}{\numberline {14}Constant of integration}{13}
\contentsline {section}{\numberline {15}Continuously differentiable}{14}
\contentsline {section}{\numberline {16}Contraction mapping}{14}
\contentsline {section}{\numberline {17}Curve}{14}
\contentsline {section}{\numberline {18}Derivation of the determining equation}{15}
\contentsline {section}{\numberline {19}Determining equation}{17}
\contentsline {section}{\numberline {20}Diagonalization}{17}
\contentsline {section}{\numberline {21}Differential form of an ODE}{17}
\contentsline {section}{\numberline {22}Differential properties of flows}{17}
\contentsline {section}{\numberline {23}Directional derivative}{17}
\contentsline {section}{\numberline {24}Distance between functions}{18}
\contentsline {section}{\numberline {25}Existence of flows}{18}
\contentsline {section}{\numberline {26}Existence of integrating factor}{18}
\contentsline {section}{\numberline {27}Existence theorem for IVP}{19}
\contentsline {section}{\numberline {28}Exponential of a matrix}{20}
\contentsline {section}{\numberline {29}Exponential of an irreducible block}{20}
\contentsline {section}{\numberline {30}Extended eigenspace}{21}
\contentsline {section}{\numberline {31}Flow of a linear system}{21}
\contentsline {section}{\numberline {32}Homogeneous equation}{21}
\contentsline {section}{\numberline {33}Homogeneous form of the determining equation}{22}
\contentsline {section}{\numberline {34}Implicit function theorem}{23}
\contentsline {section}{\numberline {35}Implicit solution}{24}
\contentsline {section}{\numberline {36}Infinitesimal symmetry}{24}
\contentsline {section}{\numberline {37}Integral curve}{25}
\contentsline {section}{\numberline {38}Integrating factor}{25}
\contentsline {section}{\numberline {39}Invariant subspace}{25}
\contentsline {section}{\numberline {40}Inverse function theorem}{26}
\contentsline {section}{\numberline {41}Irreducible nilpotent transformation}{26}
\contentsline {section}{\numberline {42}Iterative definition of sine and cosine}{27}
\contentsline {section}{\numberline {43}Jacobian}{27}
\contentsline {section}{\numberline {44}Jordan canonical form}{28}
\contentsline {section}{\numberline {45}Limit point}{28}
\contentsline {section}{\numberline {46}Limit point}{28}
\contentsline {section}{\numberline {47}Linear ODE}{29}
\contentsline {section}{\numberline {48}Linear system}{30}
\contentsline {section}{\numberline {49}Lipschitz}{30}
\contentsline {section}{\numberline {50}Method of integrating factors}{30}
\contentsline {section}{\numberline {51}Method of standard coordinates}{31}
\contentsline {section}{\numberline {52}Metric space}{32}
\contentsline {section}{\numberline {53}Nilpotent transformation}{32}
\contentsline {section}{\numberline {54}Picard iteration}{33}
\contentsline {section}{\numberline {55}Proof completeness wrt uniform convergence}{33}
\contentsline {section}{\numberline {56}Proof of existence of flows}{34}
\contentsline {section}{\numberline {57}Proof of rectification theorem}{34}
\contentsline {section}{\numberline {58}Proof of the Banach fixed point theorem}{35}
\contentsline {section}{\numberline {59}Proof of the existence theorem for IVPs}{35}
\contentsline {section}{\numberline {60}Push forward}{36}
\contentsline {section}{\numberline {61}Rectification theorem}{36}
\contentsline {section}{\numberline {62}Reducible transformation}{36}
\contentsline {section}{\numberline {63}Relation between symmetries and integrating factors}{37}
\contentsline {section}{\numberline {64}Relation between symmetries and solution curves}{37}
\contentsline {section}{\numberline {65}Separation of variables}{39}
\contentsline {section}{\numberline {66}Slope evolution formula}{40}
\contentsline {section}{\numberline {67}Slope transformation formula}{41}
\contentsline {section}{\numberline {68}Solution curve}{41}
\contentsline {section}{\numberline {69}Solution of the Bernoulli equation using the method of integrating factors}{42}
\contentsline {section}{\numberline {70}Supremum}{43}
\contentsline {section}{\numberline {71}Symmetries of homogeneous equations}{43}
\contentsline {section}{\numberline {72}Symmetries of linear equations}{44}
\contentsline {section}{\numberline {73}Symmetries of separable equations}{44}
\contentsline {section}{\numberline {74}Symmetry}{45}
\contentsline {section}{\numberline {75}The relation between limits and limit points}{46}
\contentsline {section}{\numberline {76}Transformation}{46}
\contentsline {section}{\numberline {77}Trivial symmetry}{46}
\contentsline {section}{\numberline {78}Uniform convergence}{47}
\contentsline {section}{\numberline {79}Vector field}{47}
\setlength\parskip{0pt}
\contentsline {section}{\numberline {80}Foundations assignment}{47}
\contentsline {subsection}{\numberline {80.1}Foundations problem 2a}{47}
\contentsline {subsection}{\numberline {80.2}Foundations problem 2b}{48}
\contentsline {subsection}{\numberline {80.3}Foundations problem 3a}{48}
\contentsline {subsection}{\numberline {80.4}Foundations problem 3b}{48}
\contentsline {subsection}{\numberline {80.5}Foundations problem 3c}{49}
\contentsline {subsection}{\numberline {80.6}Foundations problem 4a}{49}
\contentsline {subsection}{\numberline {80.7}Foundations problem 4b}{50}
\contentsline {subsection}{\numberline {80.8}Foundations problem 4c}{50}
\contentsline {subsection}{\numberline {80.9}Foundations problems 5a 5b}{51}
\contentsline {subsection}{\numberline {80.10}Foundations problem 5c}{53}
\contentsline {subsection}{\numberline {80.11}Foundations problem 5d}{53}
\contentsline {subsection}{\numberline {80.12}Foundations problem 5e}{53}
\contentsline {subsection}{\numberline {80.13}Foundations problem 5f}{54}
\contentsline {subsection}{\numberline {80.14}Foundations problem 5g}{54}
\contentsline {subsection}{\numberline {80.15}Foundations problem 5h}{55}
\contentsline {subsection}{\numberline {80.16}Foundations problem 5i}{55}
\contentsline {subsection}{\numberline {80.17}Foundations problems 5j}{56}
\contentsline {subsection}{\numberline {80.18}Foundations problem 6a}{56}
\contentsline {subsection}{\numberline {80.19}Foundations problem 6d}{56}
\contentsline {subsection}{\numberline {80.20}Foundations problems 6b 6c}{57}
\contentsline {subsection}{\numberline {80.21}Foundations problems 6e}{58}
\contentsline {section}{\numberline {81}Flows assignment}{59}
\contentsline {subsection}{\numberline {81.1}Flows problem 2a}{59}
\contentsline {subsection}{\numberline {81.2}Flows problem 2b}{60}
\contentsline {subsection}{\numberline {81.3}Flows problem 3}{61}
\contentsline {subsection}{\numberline {81.4}Flows problem 4}{62}
\contentsline {subsection}{\numberline {81.5}Flows problem 5}{62}
\contentsline {subsection}{\numberline {81.6}Flows problem 6a}{63}
\contentsline {subsection}{\numberline {81.7}Flows problem 6b}{63}
\contentsline {subsection}{\numberline {81.8}Flows problem 7}{64}
\contentsline {subsection}{\numberline {81.9}Flows problem 8}{65}
\contentsline {subsection}{\numberline {81.10}Flows problem 9a}{65}
\contentsline {subsection}{\numberline {81.11}Flows problem 9b}{67}
\contentsline {section}{\numberline {82}Scalar equation problems.}{68}
\contentsline {subsection}{\numberline {82.1}Scalar equations problem 1}{68}
\contentsline {subsection}{\numberline {82.2}Scalar equations problem 2}{69}
\contentsline {subsection}{\numberline {82.3}Scalar equations problem 3}{69}
\contentsline {subsection}{\numberline {82.4}Scalar equations problem 4}{70}
\contentsline {subsection}{\numberline {82.5}Scalar equations problem 5}{71}
\contentsline {section}{\numberline {83}Linear algebra problems.}{71}
\contentsline {subsection}{\numberline {83.1}linear algebra problem 1}{71}
\contentsline {subsection}{\numberline {83.2}linear algebra problem 2}{72}
\contentsline {subsection}{\numberline {83.3}linear algebra problem 3}{73}
\contentsline {subsection}{\numberline {83.4}linear algebra problem 4}{73}
\contentsline {subsection}{\numberline {83.5}linear algebra problem 5}{73}
\contentsline {subsection}{\numberline {83.6}linear algebra problem 6}{74}
}

\section{Math 5190 collaborative course notes: representative entries}
\label{NotesExamples}

\small
\setlength\parindent{0pt}
\textbf{58 \hskip 1cm Proof of the Banach fixed point theorem}\\
\textbf{Theorem}~~Let $T\colon X \to X$ be a contraction transformation of a complete
 metric space.Then there exists a unique fixed point $\hat {x}\in X$ i.e. $T(\hat{x})=
\hat{x}$.

\textbf{Proof}~~Choose a point in $X$ and call that point $x_1$.Set $$x_2=Tx_1,x_3=Tx_2,
\dots,x_{n+1}=Tx_n.$$
Set $$k=d(x_1,x_2).$$By hypothesis, $$d(x_2,x_3)\leq qk,0<q<1.$$
Continuously,we get $$d(x_n,x_{n+1})\leq kq^{n-1}.$$

Thus, 
\begin{align*}
d(x_1,x_{n+1})&\leq d(x_1,x_2)+d(x_2,x_3)+\dots+d(x_n,x_{n+1})\\
&\leq k+kq+\dots+ kq^{n-1}\\
&= k\frac{1-q^n}{1-q}.
\end{align*}

More generally, 
\begin{eqnarray*}
d(x_m,x_{n+m})&\leq& kq^{m-1}\frac{1-q^n}{1-q}\\ &\leq& \frac{dq^{m-1}}
{1-q}.  
\end{eqnarray*}
So as $m\to \infty,d(x_m,x_{n+m})\to 0$.Since $X$ is assumed to be complete,there exists a
limit of the sequence $x_n$ call that the limit $\hat x$.
Note that for all $n=1,2,\dots$,
$$d(\hat x,T\hat x)\leq d(\hat x,x_n)+d(x_n,x_{n+1})+d(x_{n+1},T\hat x) \leq (1+q)d(\hat x,
x_n)+kq^{n-1}.$$
As $n\to \infty $,the right hand side $\to 0$.
Thus, $d(\hat x,T\hat x)=0$,Therefore,$T\hat x=\hat x$.
If $\hat x,\hat y$ are such that $T\hat x=\hat x$ and $T\hat y=\hat y$,then $d(\hat x,
\hat y=d(T\hat x,T\hat y)\leq qd(\hat x,\hat y),q<1$
Thus,$d(\hat x,\hat y)=0$.Therefore, $\hat x=\hat y.$
\bigskip 

\textbf{74\hskip 1cm Symmetry}

A symmetry of a scalar ODE $\frac{dy}{dx}=\omega(x,y)$ is a transformation

$$
\begin{pmatrix}
\hat{x}\\
\hat{y}
\end{pmatrix}
=
\begin{pmatrix}
f(x,y)\\
g(x,y)
\end{pmatrix}
$$

\noindent
such that

$$
\omega(\hat{x},\hat{y})=\frac{\hat{y}_x+\omega(x,y)\hat{y}_y}{\hat{x}_x+\omega(x,y)\hat{x}_y}. 
$$

\noindent
Example 1:

\noindent
Consider the ODE

$$
\frac{dy}{dx}=-\frac{x}{y}.
$$
\noindent
We claim that 

$$
\begin{pmatrix}
\hat{x}\\
\hat{y}
\end{pmatrix}
=
\begin{pmatrix}
\cos(t) & -\sin(t)\\
\sin(t) & \cos(t)
\end{pmatrix}
\begin{pmatrix}
x\\
y
\end{pmatrix}
$$

\noindent
is a symmetry of the ODE

\begin{align*}
\frac{d\hat{y}}{d\hat{x}}&=\frac{\sin(t)+(-x/y)\cos(t)}
{\cos(t)+(-x/y)(-\sin(t))}\\
&=\frac{-x\cos(t)+y\sin(t)}{x\sin(t)+y\cos(t)}\\
&=-\frac{\hat{x}}{\hat{y}}\\
&=\omega(x,y).
\end{align*}

\noindent
Therefore the transformation is a symmetry of the ODE.

\noindent
Example 2:

\noindent
Consider the ODE

$$
\frac{dy}{dx}=x+y.
$$
\noindent
We claim that the transformation  

$$
\begin{pmatrix}
\hat{x}\\
\hat{y}
\end{pmatrix}
=
\begin{pmatrix}
x\\
y+ke^x
\end{pmatrix},
$$

\noindent
where $k$ is a constant, is a symmetry of the ODE.

\noindent
So we check:

\begin{align*}
\frac{d\hat{y}}{d\hat{x}}&=\frac{ke^x+(x+y)}{1}\\
&=x+y+ke^x\\
&=\hat{x}+\hat{y}\\
&=\omega(\hat{x},\hat{y}).
\end{align*}

\noindent
Therefore the transformation is a symmetry of the ODE.
\bigskip

\textbf{81.9\hskip 1cm Flows problem 9a}\\
What is the flow  corresponding to the following differential equation:
$$\frac{dy}{dx}  = -\frac{x}{y}.$$

Letting $\Phi_t(x,y)=\Phi(x,y,t)$ denote
the flow mapping, verify the following:

\begin{align*}
\Phi_0(x,y) &= (x,y) ;\\
\Phi_\tau \circ\Phi_t &= \Phi_{\tau+t} ;\\
\frac{\partial \Phi}{\partial t} &= \dot{\Phi}\circ\Phi ;\\
\frac{\partial\Phi}{\partial t} &= J[\Phi] \dot{\Phi}.
\end{align*}

\noindent
We first autonomize the ODE:

\begin{align*}
\frac{dx}{dt} &= 1 \\
\frac{dy}{dt} &= -\frac{x}{y}.
\end{align*}

If we solve the original ODE we see that the implicit solution is $x^2+y^2=c,$ where 
$c$ is a constant (the implicit solution can be found by separation of variables).  To 
find the flow for the above ODE, we begin by letting $\hat{x}=x+t$.  Then we note that

$$
\hat{x}^2+\hat{y}^2=c.
$$

\noindent
Solving for $\hat{y}$ gives us

\begin{align*}
\hat{y} &= \sqrt{c- \hat{x}^2}\\
        &= \sqrt{(x^2+y^2)-(x+t)^2}\\
        &= \sqrt{y^2-2xt-t^2}.
\end{align*}

\noindent
Then

$$
\Phi(t,x,y)=
\begin{pmatrix}
\hat{x}\\
\hat{y}
\end{pmatrix}
=
\begin{pmatrix}
x+t\\
\sqrt{y^2-2xt-t^2}
\end{pmatrix}.
$$

\noindent
The first condition is satisfied easily:

$$
\Phi_0(x,y)=
\begin{pmatrix}
x\\
y
\end{pmatrix}.
$$

Next we verify the condition $\Phi_t\circ\Phi_\tau = \Phi_{t+\tau}$.  Let
$$
\Phi_t=
\begin{pmatrix}
\hat{x}\\
\hat{y}
\end{pmatrix}
=
\begin{pmatrix}
x+t\\
\sqrt{y^2-2xt-t^2}
\end{pmatrix},
$$

\noindent
and let
$$
\Phi_{\tau}=
\begin{pmatrix}
\hat{\hat{x}}\\
\hat{\hat{y}}
\end{pmatrix}
=
\begin{pmatrix}
\hat{x}+\tau\\
\sqrt{\hat{y}^2-2\hat{x} \tau- \tau^2}
\end{pmatrix}.
$$

\noindent
Then

\begin{align*}
\Phi_\tau \circ\Phi_t &=
\begin{pmatrix}
x+t\\
\sqrt{y^2-2xt-t^2-2(x+t)\tau-\tau^2}
\end{pmatrix} \\
&=
\begin{pmatrix}
x+(\tau+t)\\
\sqrt{y^2-2x(\tau+t)-(\tau+t)^2}
\end{pmatrix} 
=\Phi_{\tau+t}.
\end{align*}

\noindent
Now consider 

$$
\frac{\partial \Phi}{\partial t}=
\begin{pmatrix}
1\\
-\frac{x+t}{\sqrt{y^2-2xt-t^2}}
\end{pmatrix}.
$$

\noindent
We test the two remaining conditions.  First we check to see if 
$\frac{\partial \Phi}{\partial t} = \dot{\Phi}\circ\Phi$.

\noindent
Note that

$$
\dot{\Phi}=
\begin{pmatrix}
1\\
-\frac{x+t}{y^2-2xt-t^2}
\end{pmatrix}_{t=0}
=
\begin{pmatrix}
1\\
-\frac{x}{y}
\end{pmatrix}.
$$
\noindent
Then
\begin{align*}
\dot{\Phi}\circ\Phi &= \dot{\Phi}(\hat{x},\hat{y})\\
&=
\begin{pmatrix}
1\\
-\frac{\hat{x}}{\hat{y}}
\end{pmatrix}\\
&=
\begin{pmatrix}
1\\
-\frac{x+t}{\sqrt{y^2-2xt-t^2}}
\end{pmatrix}.
\end{align*}

\noindent
Therefore $\frac{\partial \Phi}{\partial t} = \dot{\Phi}\circ\Phi$.  Finally we test the 
condition $\frac{\partial\Phi}{\partial t} = J[\Phi] \dot{\Phi}$.

\begin{align*}
J[\Phi]\dot{\Phi} &=
\begin{pmatrix}
1 & 0\\
-\frac{t}{\sqrt{y^2-2xt-t^2}} & \frac{2y}{\sqrt{y^2-2xt-t^2}}
\end{pmatrix}
\begin{pmatrix}
1\\
-\frac{x}{y}
\end{pmatrix}
\\
&= 
\begin{pmatrix}
1\\
-\frac{x+t}{\sqrt{y^2-2xt-t^2}}
\end{pmatrix}
= \frac{\partial \Phi}{\partial t}.
\end{align*}

\noindent
Therefore our four conditions stated in the problem above are satisfied by the flow we 
had determined, $\Phi$.

\end{document}